\documentclass{xray_symp}
\usepackage{graphics}
\def\ros{{\sl ROSAT }}

\def\G{$\Gamma_{\rm x}$ }
\def\approxgt{\mathrel{\hbox{\rlap{\lower.55ex \hbox {$\sim$}}
        \kern-.3em \raise.4ex \hbox{$>$}}}}
\begin{document}
\title{The giant-amplitude X-ray outburst in NGC\,5905 \\ 
-- a tidal disruption event ? }
\author{Stefanie Komossa\inst{1}, Norbert Bade\inst{2} } 
\institute{Max--Planck--Institut f\"ur extraterrestrische Physik,
 Giessenbachstra{\ss}e, 85740 Garching, Germany
\and Hamburger Sternwarte, Gojenbergsweg 112, 21029 Hamburg, Germany} 
\authorrunning{St. Komossa, N. Bade}
\titlerunning{~The giant X-ray outburst in NGC\,5905 -- a tidal disruption event ?}
\maketitle

\begin{abstract}

NGC 5905 is one of the very few galaxies{\footnote {the other three
objects are the Seyfert galaxies IC\,3599 (Brandt et al. 1995, Grupe et al. 1995a)
E\,1615+061 (Piro et al. 1988) and WPVS007 (Grupe et al. 1995b)
}}
that underwent a giant X-ray outburst,
with a change in photon countrate of a factor $\sim$100.
The outburst spectrum is
both, very soft and luminous (Bade, Komossa \& Dahlem 1996).
Remarkably,  the optical pre-outburst emission line spectrum
of NGC 5905 is that of an HII-type galaxy, i.e. does not show 
any signs of 
Seyfert activity.

One exciting explanation of the X-ray observation is that we may have witnessed
the tidal disruption of a star by a supermassive black hole (SMBH) residing in
the nucleus of this galaxy. The expected flare of electromagnetic
radiation being produced when the stellar debris is swallowed by the black hole
was proposed by Rees (1988) as a means of tracing SMBHs in nearby non-active galaxies.

In the present work, we discuss this and other possible outburst scenarios
through an analysis of all \ros PSPC and HRI X-ray observations of NGC 5905.
We also present {\em simultaneous} and long-term optical photometry of NGC 5905 
as well as
follow-up optical spectroscopy. 

\end{abstract}

\section{X-ray observations}

A giant X-ray outburst of NGC 5905 was discovered during the \ros 
(Tr\"umper 1983) survey observation
(Bade, Komossa \& Dahlem 
1996).  
NGC 5905 showed a high countrate during this first \ros observation. The countrate
then declined by at least a factor of several within months and was down by 
a factor $\sim$100 two years later. 

The X-ray spectrum during the outburst was very soft (with photon index \G $\simeq -4.0$ when
fit by a powerlaw). During quiescence, the spectrum is flatter
(\G $\simeq -2.4$). 
The outburst luminosity is of the order of
$L_{\rm x} \approxgt$ several $\times~10^{42}$ erg/s; much {\em higher} than 
observed in {\em non-active} spiral galaxies (e.g., Fabbiano 1989, Vogler 1997). 

New HRI data were taken in 1996 with an exposure time
of 76 ksec. These show a further decline
by a factor $\sim$2 in flux with respect to the last PSPC observation.
The long-term X-ray lightcurve is displayed in Fig. 1. 

  \begin{figure*} 
\vspace*{-1.8cm}
 \resizebox{\hsize}{!}{\includegraphics{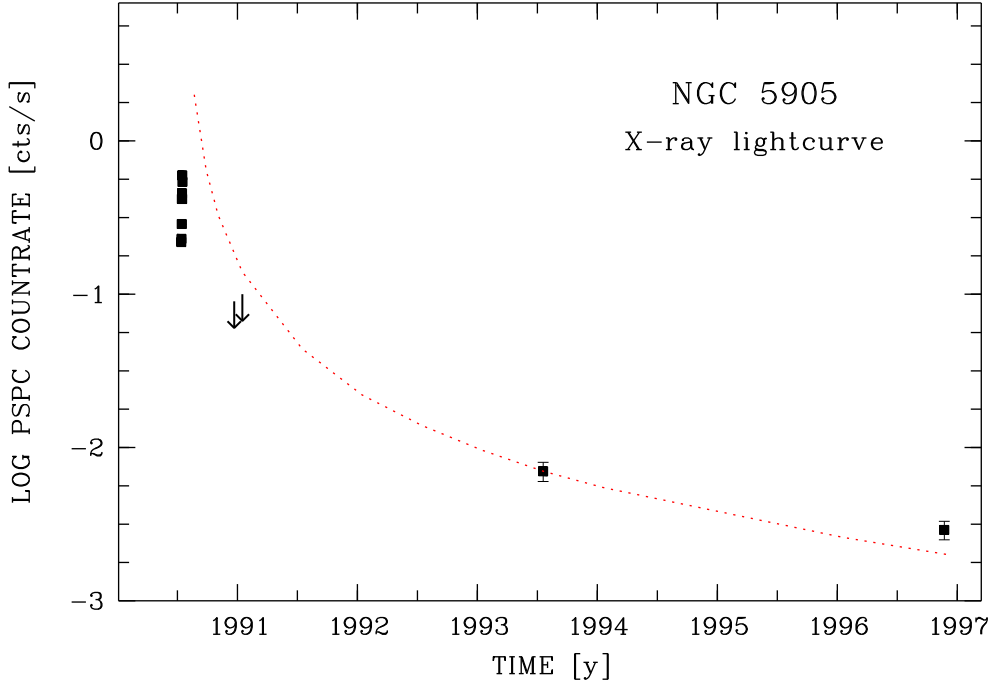}}
\vspace*{-1.4cm}
 \caption[comp]{X-ray lightcurve of NGC 5905 observed with the \ros PSPC and HRI.
The dotted curve follows the relation $CR = 0.044 (t - t_{\rm o})^{-5/3}$
with $t_{\rm o}$=1990.54. A time dependence of $\propto t^{-5/3}$ is predicted
in the model of tidal disruption of a star (Rees 1988, 1990). 
} 
\label{comp}
\end{figure*}

\section{Optical observations}

Optical observations are an important supplement to the X-ray data.  
The two most important questions are (i) is there a simultaneous optical outburst,
and (ii) does the optical spectrum of NGC 5905 show any signs of Seyfert activity ?

\subsection{Photometry} 
We have used photographic plates taken at {\sl Sternwarte Sonneberg} 
to produce a long-term optical lightcurve (between years 1962 and 1995) of the nucleus of NGC 5905 
and searched the plates  taken quasi-{\em simultaneously} to the
X-ray outburst for a correlated optical outburst. A change
in brightness of 5$^{\rm m}$ is expected if the amplitude at optical
wavelengths is the same as in X-rays.  

\begin{figure} 
 \resizebox{\hsize}{!}{\includegraphics{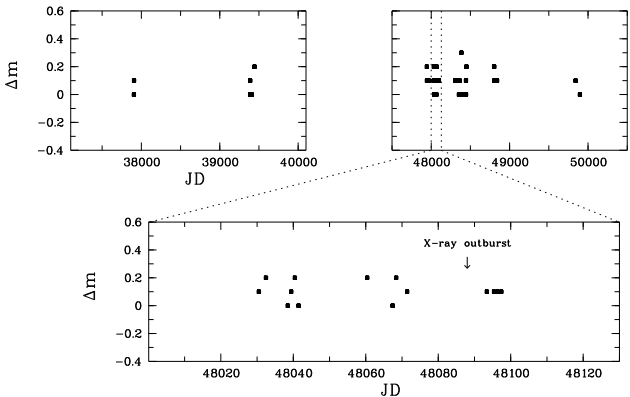}}
 \caption[n5905_sonne]{Optical lightcurve of the nucleus of NGC 5905, based on 
photographic plates taken at {\sl Sternwarte Sonneberg}.
The abscissa gives the Julian date in JD-2400000 (the data points bracket the 
time interval 1962 -- 1995).  
The upper panel shows the longterm lightcurve, the lower one the lightcurve in the months
around the X-ray outburst. 
Within the error of about 0.2$^{\rm m}$
the optical brightness is constant. 
}
\label{n5905_sonne}
\end{figure}

The optical brightness of NGC 5905 is found to be {\em constant} within 
the errors on long terms as well as near the X-ray outburst (Fig. \ref{n5905_sonne}), 
placing strong constraints on outburst scenarios.
%

\subsection{Spectroscopy}

\begin{figure*} 
\vspace{-0.5cm}
 \resizebox{\hsize}{!}{\includegraphics{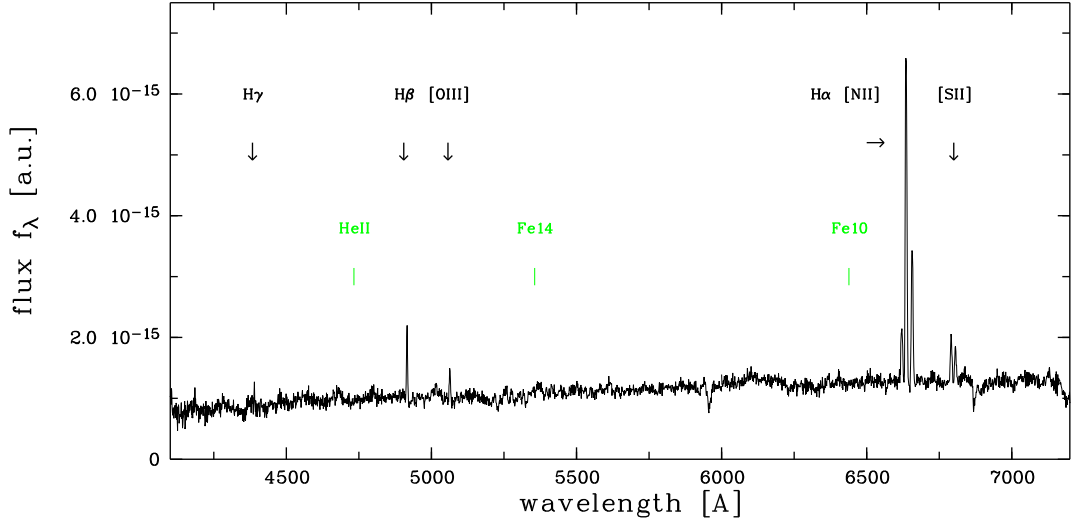}} 
\vspace{-3.7cm}
 \caption[n5905_opt]{Optical spectrum of the nucleus of NGC 5905 taken
about 6 years after the X-ray outburst.   
The arrows mark the detected emission lines. 
The weakness of [OIII]$\lambda$5007 compared to H$\beta$ classifies the galaxy 
as HII-type.
The positions of some high-ionization lines, not detected in the spectrum,
are marked as well.   
}
\label{n5905_opt}
\end{figure*}

We have taken high-resolution post-outburst optical spectra (about 6 years after the X-ray outburst)  
at the 3.5\,m telescope of Calar Alto (Fig. \ref{n5905_opt}).
These show the same spectral characteristics as the pre-outburst spectrum of Ho et al. (1995)
and classify the galaxy as HII-type. 
If a high-ionization emission-line component was present during outburst, it had already 
declined below detectability. We carefully searched for other signs of {\em permanent} 
Seyfert activity --
none are revealed. This is another important constraint for outburst models
and, at present, makes NGC 5905 the only non-active galaxy among the X-ray outbursting 
ones.  

\section{Outburst scenarios}

The major characteristics to be explained by outburst models are
\begin{itemize} 
\item 
short duration, 
extremely soft spectrum, giant amplitude (a factor $\approxgt$ 100)  
and huge peak luminosity 
\mbox{($L_{\rm x} \approxgt 10^{42-43}$ erg/s)} of the X-ray outburst 
\item 
no simultaneous change in optical brightness 
\item
the optical spectrum is that of an HII-type galaxy with {\em no} signs of {\em permanent}
Seyfert activity  
\end{itemize}

\noindent 
We in turn discuss several outburst scenarios (for details see Komossa \& Bade 1998). 

\subsection {Supernova in dense medium}

The outburst X-ray luminosity by far exceeds that of individual
supernovae (those observed in X-rays
range between $L_{\rm x} \approx 10^{35}$ and a few $10^{40}$ erg/s;
e.g., Schlegel 1995, Immler \& Pietsch 1998). 

The possibility of `buried' supernovae (SN) in {\em dense} molecular gas
was studied by Shull (1980) and
Wheeler et al. (1980).
In this scenario, X-ray emission originates from the shock, produced by the expansion of the SN ejecta
into the ambient interstellar gas of high density.
Since high luminosities can be reached this way, and the evolutionary time
is considerably speeded up, an SN in a dense
medium may be an explanation for the observed X-ray outburst in NGC 5905.

Assuming the observed outburst luminosity of NGC 5905 to be the peak luminosity
and using the analytical estimates of Shull and Wheeler et al.
for the evolution of temperature, radius and luminosity of the shock,
\begin {equation}
T = E_{51}^{0.14} n_4^{0.27} (t/t_{\rm rad})^{-10/7} (2.7\, 10^7\, \rm K)~~,
\end {equation}
\begin {equation}
R= E_{51}^{0.29} n_4^{-0.43} (t/t_{\rm rad})^{2/7} (0.29\, \rm {pc})~~,
\end {equation}
\begin {equation}
L= E_{51}^{0.78} n_4^{0.56} (t/t_{\rm rad})^{-11/7} (9.8\, 10^{39}\, \rm {erg/s})~~,
\end {equation}
results in a density of the ambient medium of $n \simeq 10^6 \rm {cm}^{-3}$
but is inconsistent with the observed softness of the spectrum:
The expected temperature is $T \approx 10^8$ K, compared to the observed one of
$T \approx 10^6$ K.
Therefore, an SN in dense medium is an unlikely explanation of the observed X-ray outburst.
Additionally, fine-tuning in the column density of the surrounding medium
would be required in this scenario in order to prevent the SNR from being completely self-absorbed.

\subsection {Gravitational lensing event}

Large magnification factors of the source brightness can be reached by
gravitational lensing (GL) if the observer, lens, and source ly nearly exactly on a 
line. Lensing was, e.g., discussed as a possibility to 
explain the observed variability in BL Lac objects (e.g. Ostriker \& Vietri 1985).      
Being independent of photon energy, GL predicts the same magnification factor
at optical wavelengths as in X-rays.   
Given the non-detection of optical variability simultaneous to the X-ray outburst the GL scenario
is very unlikely. 

\subsection {X-ray afterglow of a Gamma-Ray Burst}

As to the question of whether the observations may have represented
the X-ray afterglow of a Gamma-Ray Burst (GRB), we note that no GRB has been detected
in the time around the X-ray outburst (July 12--15, 1990) of NGC\,5905
(the 2 reported GRBs nearest in time, of 
July 8, 1990, have different positions; Castro-Tirado 1994 on the basis of Granat/WATCH data).
The possibility remains that the GRB escaped detection or that it was not beamed 
towards us and only the isotropic afterglow was seen.

\subsection {Warm-absorbed hidden Seyfert nucleus}

In this
scenario, NGC\,5905 would host a Seyfert nucleus
which is intrinsically slightly variable and usually hidden behind a 
large cold column of absorbing
gas. The nucleus gets visible in X-rays during its flux high-states by {\em ionizing}
the originally {\em cold} column
of surrounding gas that becomes a {\em warm} absorber, which is transparent
to soft X-rays and could also account for the softness of the outburst spectrum
(Komossa \& Fink 1997a).
We find that the model of a (dust-free) warm absorber, calculated with Ferland's (1993)
code {\em Cloudy}, 
successfully fits the X-ray spectrum with an ionization parameter log $U \simeq$ 0 and
a column density of the warm material log $N_{\rm w} \simeq$ 22.8.
Since there is no evidence for Seyfert activity in the optical spectrum,
the nucleus must be hidden.
Mixing dust with the warm gas
could hide the Seyfert nucleus completely.
However, dust with Galactic ISM properties mixed with the warm gas
strongly influences the X-ray absorption structure (e.g. Komossa \& Fink 1997b,
and these proceedings) and
a fit of a {\em dusty} warm absorber to the X-ray spectrum 
can only be achieved if we strongly fine-tune the
dust properties and the dust depletion factor.

\subsection {Accretion-disk instability}

If a massive BH {\em with an accretion disk}
exists in the center of NGC 5905, it usually has to accrete with low
accretion rate or radiate with low efficiency, to account for the
comparatively low X-ray luminosity of NGC 5905 in quiescence.
An accretion disk instability may then provide an explanation for the observed X-ray outburst.
Thermally unstable slim accretion disks were studied by Honma et al. (1991),
who find the disk to exhibit burst-like oscillations for the case of the standard $\alpha$
viscosity description and for certain values of accretion rate.

Using the estimate for the duration of the  high-luminosity state
(Honma et al.; their Eq. 4.8),
and a duration of the outburst of less than 5 months
(the time difference between the first two observations of NGC 5905),
a central black hole of mass in the range $\sim 10^4 - 10^5 M_{\odot}$
could account for the observations.
The burst-like oscillations are found by Honma et al. only for certain values of the initial accretion
rate.
A more detailed quantitative comparison with the observed outburst in NGC\,5905 is difficult,
since the behavior of the disk is quite model dependent, and further detailed modeling
would be needed.

\subsection {Tidal disruption of a star}

The idea of tidal disruption of stars by a supermassive black hole (SMBH)
was originally studied as a possibility to fuel AGN (Hills 1975), but was dismissed later.
Peterson \& Ferland (1986) suggested this mechanism as possible explanation for
the transient brightening of the HeII line observed in a Seyfert galaxy.
Tidal disruption was invoked by Eracleous et al. (1995) in a model
to explain the UV properties of LINERs, and was suggested as possible outburst
mechanism for IC\,3599 (Brandt et al. 1995, Grupe et al. 1995).  

Rees (1988, 1990) proposed to use individual such events
as tracers of SMBHs in nearby {\em non-active} galaxies.
The debris of the disrupted star is accreted by the BH.
This produces a flare, lasting of the order
of months, with the peak luminosity in the opt-UV, EUV or soft X-ray spectral region.

The luminosity emitted if the BH is accreting at its Eddington luminosity
can be estimated by $ L_{\rm edd} \simeq 1.3 \times 10^{38} M/M_{\odot}$ erg/s.
In case of NGC 5905, a BH mass of at least $\sim 10^{4-5}$ M$_{\odot}$ would be
required to 
produce the observed $L_{\rm x}$, and a higher mass if $L_{\rm x}$ was not observed 
at its peak value. 

The decline in luminosity after the maximum in the tidal disruption scenario 
scales as $L \propto t^{-{5\over 3}}$ 
(Rees 1990). The observed long-term X-ray lightcurve of NGC 5905 is given in Fig. 1.   

We favour the scenario of tidal disruption of a star because it can account
for the high outburst luminosity,
seems to require least fine-tuning, and the long-term X-ray lightcurve
shows a continuous fading of the source over the whole measured time
interval.

\section{Search for other strongly variable objects}

We performed a search for further cases of strong X-ray variability (Komossa 1997)
using the sample of nearby galaxies of Ho et al. (1995) and \ros survey
and archived pointed observations.
We do not find another object with a factor $\sim$100 amplitude,
but several sources are discovered to be strongly variable (with maximal factors ranging
between 10 and 20 in the mean countrates):
NGC\,3227, NGC\,4051, and NGC\,3516. Another case of this amplitude, NGC\,3628,
was reported by Dahlem et al. (1995).

\section {Summarizing conclusions}
 
The most likely scenario to explain the X-ray outburst in NGC 5905
seems to be tidal disruption of a star by a central SMBH; a scenario  
proposed by Rees (1988) as a tracer of SMBHs in nearby galaxies.
We caution, though, that many theoretical details of this process are still rather
unclear. 

High-resolution optical spectroscopy of NGC 5905 does not reveal any signs of
Seyfert activity. At present, this makes NGC 5905 the {\em only} non-active galaxy
among the X-ray outbursting ones (Komossa \& Bade 1998).   

The X-ray outburst in this HII galaxy then lends further support to the scenario that
{\em all} galaxies passed through an active phase (instead of just a few), leaving
unfed SMBHs in their centers. 

\begin{acknowledgements}
It is a pleasure to thank the {\sl Sternwarte
Sonneberg} for the kind
hospitality and G. Richter additionally for 
valuable help in the assessment of the photographic plates. 
St.K. acknowledges support from the Verbundforschung under grant No. 50\,OR\,93065.
\end{acknowledgements}

\end{document}